\documentclass[letterpaper]{article} 
\usepackage{aaai24}  
\usepackage{times}  
\usepackage{helvet}  
\usepackage{courier}  
\usepackage[hyphens]{url}  
\usepackage{graphicx} 
\usepackage{natbib}  
\usepackage{caption} 
\DeclareCaptionStyle{ruled}{labelfont=normalfont,labelsep=colon,strut=off} 
\usepackage{algorithm}
\usepackage{algorithmic}

\usepackage{xcolor}

\usepackage{newfloat}
\usepackage{listings}
\floatstyle{ruled}
\newfloat{listing}{tb}{lst}{}
\floatname{listing}{Listing}
\title{IsamasRed: A Public Dataset Tracking Reddit Discussions \\on Israel-Hamas Conflict}
\author{
    Kai Chen\textsuperscript{\rm 1,2},
    Zihao He\textsuperscript{\rm 1,2},
    Keith Burghardt\textsuperscript{\rm 1},
    Jingxin Zhang\textsuperscript{\rm 2},
    Kristina Lerman\textsuperscript{\rm 1,2}
}
\affiliations{
    \textsuperscript{\rm 1}Information Sciences Institute, University of Southern California\\
    \textsuperscript{\rm 2}Department of Computer Science, University of Southern California\\

    kchen035@usc.edu, 
    zihaoh@usc.edu, 
    keithab@isi.edu, 
    jzhang78@usc.edu, 
    lerman@isi.edu
    \\
}

\usepackage{bibentry}

\begin{document}

\maketitle

\begin{abstract}

The conflict between Israel and Palestinians significantly escalated after the October 7, 2023 Hamas attack, capturing global attention. To understand the public discourse on this conflict, we present a meticulously compiled dataset--IsamasRed--comprising nearly 400,000 conversations and over 8 million comments from Reddit, spanning from August 2023 to November 2023. We introduce an innovative keyword extraction framework leveraging a large language model to effectively identify pertinent keywords, ensuring a comprehensive data collection. Our initial analysis on the dataset, examining topics, controversy, emotional and moral language trends over time, highlights the emotionally charged and complex nature of the discourse. This dataset aims to enrich the understanding of online discussions, shedding light on the complex interplay between ideology, sentiment, and community engagement in digital spaces.

\end{abstract}

\section{Introduction}
On October 7th, 2023, Hamas militants launched an wide-ranging assault, firing thousands of rockets from Gaza towards Israel and infiltrating towns and villages in southern Israel. The attack resulted in unprecedented Israeli casualties, including more than a thousand deaths and 240 hostages.  In response, the Israeli government, led by Prime Minister Netanyahu, launched a counter attack that  resulted in massive casualties in Gaza. The ongoing war in the Middle East threatens to further destabilize the region and expand the crisis far beyond the countries involved. 

The international response to the Israel-Hamas conflict was swift and divisive. The mass demonstrations and counter demonstrations across the world exposed long-simmering  grievances and cultural fault lines. Loud voices exacerbated deep divisions and generational divides about the legitimacy of Israel vs the legitimacy of Palestine, right to self-defense and self-determination, Zionism vs Palestinian rights, antisemitism vs Islamophobia, etc \cite{samuel2023israel}. This debate has grown especially contentious in the U.S., where it is set against the political backdrop of a presidential campaign and ongoing culture wars.


To better understand public discourse about the Israel-Hamas conflict and how it played out on social media, we aim to collect data from Reddit, a popular social media platform featuring user-generated content within topic-specific communities called subreddits \cite{anderson2015ask, datta2019extracting}. Its diversity \cite{weninger2013exploration} and active participation \cite{choi2015characterizing} make Reddit an excellent source for data collection~\cite{jamnik2019use}, particularly on geopolitical issues like the Israel-Hamas conflict. Subreddits allow for focused data gathering, providing insights into public opinion and discourse \cite{hofmann2022reddit, petruzzellis2023relation}. Conversations on Reddit, including submissions and comments, offer valuable, detailed discussions and diverse viewpoints, contributing significantly to research on complex social and political topics by providing real-time engagement and sentiment analysis \cite{park2021detecting, he2023cpl}.

When collecting a dataset of public online discussions, researchers use keywords to target data collection to  relevant discussions from among the vast volume of message posted online~\cite{chen2020tracking, zhu2022reddit, chang2023roeoverturned, chen2023tweets}. However, the keywords used for retrieving messages are usually chosen manually, in an ad hoc manner, which risks biasing data collection. To address this challenge, we introduce a framework that leverages a large language model \cite{ouyang2022training} to automate keyword extraction, significantly enhancing the identification of relevant keywords. 
Specifically, we first retrieve Wikipedia pages on the conflict using a small set of seed terms describing the topic. The model then identifies keywords pertinent to the conflict from the pages and scores them based on semantic relevance, leading to a ranked list of keywords. Our approach yields a comprehensive and optimized set of query keywords, which significantly enhances data collection and in-depth analysis of the complex dynamics involved in the conflict.

Focusing on Reddit as a primary source, we use this framework as a systematic strategy for gathering a wide array of perspectives and narratives about the Israel-Hamas conflict. In total, we collect over 400K conversations including more than 8M comments. To the best of our knowledge, ours is the first large scale dataset on the \textbf{Is}rael-H\textbf{amas} Conflict discourse on \textbf{Red}dit, and we name our dataset IsamasRed. From IsamasRed, we further identify two subsets, comprising discourse related to Zionism/Antisemitism (IsamasRed-Z) and Free Palestine/Islamophobia (IsamasRed-P) respectively.

We present a comprehensive analysis of IsamasRed, focusing on key subreddits to explore the mechanisms of public opinion formation and narrative dynamics. We systematically examine the structural and substantive aspects of user-generated content, employing quantitative methods to track engagement and qualitative approaches to interpret discourse patterns, particularly in response to significant events in the conflict.
Specifically, we first measure user endorsement of discussions through the quantity of upvotes and downvotes they have received.
We then explore the controversial nature of the conflict, employing Reddit's controversiality indicator to measure debate intensity and understand the polarizing factors that fuel online discussions.
To quantify the ethical and emotional dimension of online discourse, we use state-of-the-art transformer models \cite{devlin2018bert, chochlakis2023emotion, chochlakis2023leveraging, guo2023data} to detect moral sentiments and emotions in posts. 
This helps better understand the discourse's affective and ethical layers, offering insights into how users align themselves with various aspects of the conflict and how these alignments influence community interactions and narrative construction. Through this multifaceted analysis, our study aims to contribute a rich, contextually informed understanding of the digital discourse surrounding the Israel-Hamas conflict.

In summary, our study makes the following contributions: (1) We introduce a novel keyword extraction framework that utilizes a large language model to significantly enhance critical keyword identification through semantic and structural analysis. (2) We build a comprehensive dataset (IsamasRed) featuring 400K conversations and 8M comments from Reddit about the Israel-Hamas conflict. (3) We conduct a thorough analysis of this dataset, examining submissions, comments, and conversation threads, alongside deeper analytical dives into user endorsement, controversiality, emotions, and moral sentiments.

\section{Related Work}

\subsection{Social Media and Geopolitical Conflicts}

The Israel-Palestine Conflict dates back to the start of 20th century, with unique historical, terrirorial, and political contexts.
\citet{imtiaz2022taking} collect tweets regarding this conflict in 2021, and study the stances (pro-Palestinian, pro-Israel, or neutral) of the tweets towards the conflict, before the conflict intensified in 2023. However, few studies on social media analysis have focused on the conflict after the escalation in 2023. In this paper, we aim to fill this gap.

Another ongoing geopolitical conflict with roots in the early 20th century is the Ukraine-Russia Conflict, which escalated in 2022 with Russia's invasion of Ukraine.
Existing works have collected extensive social media tweets to facilitate research into various aspects of the conflict, such as social media influence, public engagement, content moderation, and the evolution of conflict narratives \cite{chen2023tweets, pohl2023invasion, zhu2022reddit, shi2023safer}. 
\citet{guerra2023sentiment} propose a lexicon-based unsupervised sentiment analysis method to measure \emph{hope} and \emph{fear} for the conflict using data from Reddit.
\citet{krivivcic2023analyzing} and \citet{nandurkar2023sentiment} study people's opinions about the conflict using VADER sentiment analysis on Reddit comments.
\citet{hanley2023happenstance} analyzes the coordinated misinformation campaign by Russian state media during the Ukraine invasion.

\subsection{Keyword Extraction}Keyword extraction is a fundamental task in text mining and natural language processing, serving as a cornerstone for various applications such as information retrieval, content summarization, and topic modeling \cite{firoozeh2020keyword, nasar2019textual, barde2017overview}. The evolution of keyword extraction methods has progressed from statistical approaches like Term Frequency-Inverse Document Frequency (TF-IDF) and TextRank \cite{mihalcea2004textrank} to transformer-based techniques \cite{devika2021deep}. \citet{burghardt2023socio} employ SAGE model \cite{eisenstein2011sparse} for keyword calculation in concern detection, while \citet{grootendorst2020keybert} introduce keyBERT and keyLLM for advanced keyword extraction. In this work, we propose to use large language models (LLMs) for automatic keyword extraction.

\begin{figure*}[ht]
    \centering
    \includegraphics[width=0.9\linewidth]{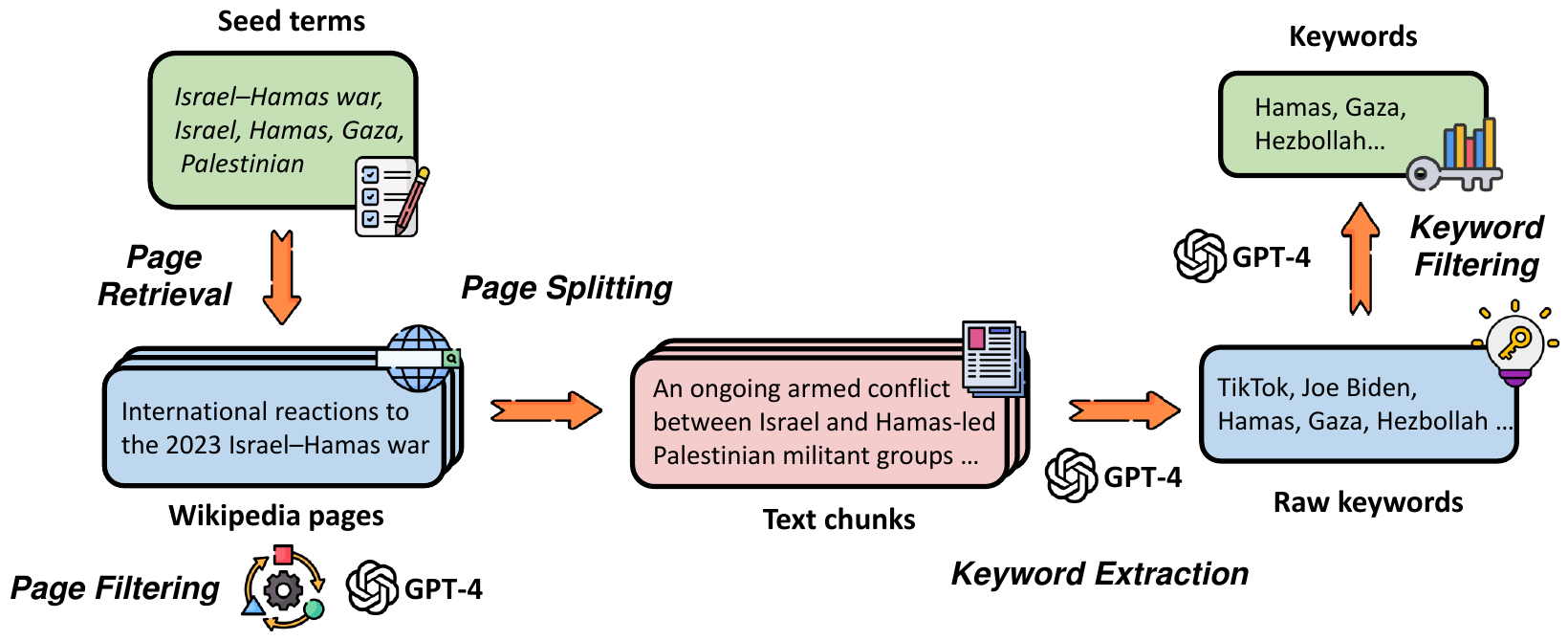}
    \caption{The framework of automated keyword extraction via LLMs. We first retrieve relevant Wikipedia pages using a small set of seed terms. Subsequently, we employ GPT-4 to filter out pages with weak relevance. Each page is then segmented into text chunks to fit the context window of the model. GPT-4 is used to identify keywords from the chunks and score them based on semantic relevance.
    Finally, the generic keywords not directly related to the topic of interest are filtered out.
    }
    \label{fig:framework}
\end{figure*}

\begin{table*}[ht]
\renewcommand{\arraystretch}{1.2}
\begin{tabular}{|l|l|}
\hline
\textbf{Category}           & \textbf{Subreddits}                           \\ \hline
Conflict-centric Subreddits     & \begin{tabular}[c]{@{}l@{}}Palestine, IsraelPalestine, AskMiddleEast, IsraelHamasWar, islam, israelexposed, \\ exmuslim, Jewish, Judaism, IsraelCrimes, Palestinian\_Violence, AntiSemitismInReddit\\ IsraelWarVideoReport, IsraelUnderAttack, Israel\_Palestine, IsraelICYMI, IsraelWar, \\ IsrealPalestineWar\_23, MuslimLounge, Muslim, Gaza, MuslimCorner, IsraelVsHamas, \\ Israel, PalestinianvsIsrael\end{tabular} \\ \hline

Conflict-inclusive Subreddits & \begin{tabular}[c]{@{}l@{}}AutoNewspaper, worldnews, news, brasilnoticias, AskReddit, Destiny, 2ndYomKi-\\ ppurWar, CombatFootage, DisneyNewsfeed, TrendingQuickTVnews, Conservative, \\ BreakingNews24hr, conspiracy, EndlessWar, PublicFreakout, politics, NoStupid-\\ Questions, SeenOnNews\_longtail, NBCauto, raceplay, FreeKarma4You, europe, \\ NoFilterNews, worldnewsvideo, TheDeprogram ...\end{tabular}        \\ \hline
\end{tabular}
\caption{The complete 25 conflict-centric subreddits and a subset of 25 conflict-inclusive subreddits.}
\label{tab:subreddits}
\end{table*}

\section{Methods}
\subsection{Automated Keyword Extraction with LLM}

Traditional methods to collect social media data on certain topics, such as the COVID-19 pandemic~\cite{chen2020tracking}, abortion rights in the US~\cite{chang2023roeoverturned}, and the Russia-Ukraine war \cite{chen2023tweets}, use manually-chosen keywords to retrieve relevant posts. However, manual selection is ad hoc and risks introducing unwanted biases into data collection.

To address this challenge, we propose a novel framework that utilizes a large language model (LLM) GPT-4 to automate keyword extraction. The overview of the framework is shown in Figure~\ref{fig:framework}. The process comprises the following steps:

\begin{itemize}
    \item \textbf{Wikipedia page retrieval.} Based on a set of seed terms (\emph{Israel–Hamas war}, \emph{Israel}, \emph{Hamas}, \emph{Palestinian}, \emph{Gaza}) relevant to the Israel-Hamas conflict, we retrieve 120 matching Wikipedia pages. Note that the seed terms are directly related to the topic of interest and does not require deliberation.
    We employ Wikipedia pages due to their comprehensive and neutral coverage of a wide array of topics \cite{he2022infusing}, including the Israel-Hamas conflict. Wikipedia's structured, referenced content and its dynamic, continually updated nature provide a balanced and extensive foundation for initial keyword generation. This approach helps mitigate potential biases and ensures that the keywords extracted by our large language model capture the nuanced and evolving nature of the conflict, thereby facilitating a more accurate and insightful collection of social media data.
    
    \item \textbf{Page filtering.} We use GPT-4 to filter out retrieved pages that are weakly related to the conflict, based on the page titles and the first 100 tokens of the pages, using prompt detailed in the Appendix. We are left with 50 pages after the filtering.
    
    \item \textbf{Page splitting.} We divide the Wikipedia pages into multiple text chunks, of up to 3,000 tokens per chunk. The step is to make sure that the text chunks do not exceed the maximum window size of GPT-4 in the following step of keyword extraction.
    
    \item \textbf{Keyword extraction.} Given a text chunk, we use GPT-4 to extract keywords from it, along with the importance scores of the keywords. The importance of each keyword is rated on a scale of 0 (least important) to 5 (most important). Please refer to the Appendix for the crafted prompt. For keywords appearing multiple times across different chunks of the same page, we merge them and average their importance values. For the same keywords appear across different pages, we sum their importance values. Ultimately, we rank the extracted keywords based on their importance values, keeping the top 200 most relevant keywords to the conflict. 

    \item \textbf{Keyword filtering.} Upon reviewing these top-ranked keywords, we note that they include terms of generic entities, such as social media platforms (\emph{TikTok}),  news organizations (\emph{CNN}, \emph{BBC}), countries (\emph{US}, \emph{China}), and political figures (\emph{Biden}). These keywords are too general for retrieving social media posts related to the conflict.
    We use GPT-4 to filter out the general keywords, resulting in a list of of 174 keywords specific to the Israel-Hamas conflict, providing a basis for subsequent data collection and analysis. The prompt for keyword filtering using GPT-4 is shown in the Appendix.
\end{itemize}

\begin{table}[ht]
\renewcommand{\arraystretch}{1.2}
\begin{tabular}{ccc}
\hline
              & \# of submissions & \# of comments \\ \hline
IsamasRed   & 412,258           & 8,089,095      \\
IsamasRed-Z & 126,107           & 2,516,114      \\
IsamasRed-P & 75,625            & 1,408,967      \\ \hline
\end{tabular}
\caption{Number of submissions and comments in IsamasRed (the complete dataset), IsamasRed-Z (the subset on Zionism/antisemitism), IsamasRed-P (the subset on Palestine/Islamophobia).}
\label{tab:data_stats}
\end{table}

\subsection{Data Collection}

We downloaded the Reddit data dump from Academic Torrent\footnote{https://academictorrents.com/}, which collects Reddit submissions using the Pushshift API \cite{baumgartner2020pushshift}. This dataset spans from August 2023 to November 2023 and encompasses all submissions (also called ``posts'') and comments across various subreddits. 

To filter the submissions related to the Israel-Hamas conflict, we start with keyword matching on the submissions. Specifically, if the title of the submission contains at least one word from the extracted keyword list, the submission is considered a match. We count the number of matched submissions for each subreddit. From the top subreddits with the most matched submissions, we select 25 \textbf{conflict-centric subreddits} (e.g., \emph{r/IsraelPalestine} and \emph{r/IsraelHamasWar}) and 75 \textbf{conflict-inclusive subreddits} (e.g., \emph{r/AutoNewspaper} and \emph{r/worldnews}). 
Conflict-centric subreddits are the ones primarily focused on the specific conflict, and conflict-inclusive subreddits are not centered on the conflict but frequently feature relevant content. We present the 25 conflict-centric subreddits and a subset of 25 conflict-inclusive subreddits in Table \ref{tab:subreddits}. For the complete set of the 75 conflict-inclusive subreddits please refer to the Appendix.
For conflict-centric subreddits, all submissions and their successor comments are collected to our dataset. For conflict-inclusive subreddits with a mix of relevant and irrelevant submissions, we only filter the matched ones and the comments. 
Table \ref{tab:data_stats} presents the statistics of our dataset \textbf{Is}rael-H\textbf{amas}-\textbf{Red}dit (IsamasRed).



\begin{figure}
    \centering
    \includegraphics[width=0.48\textwidth]{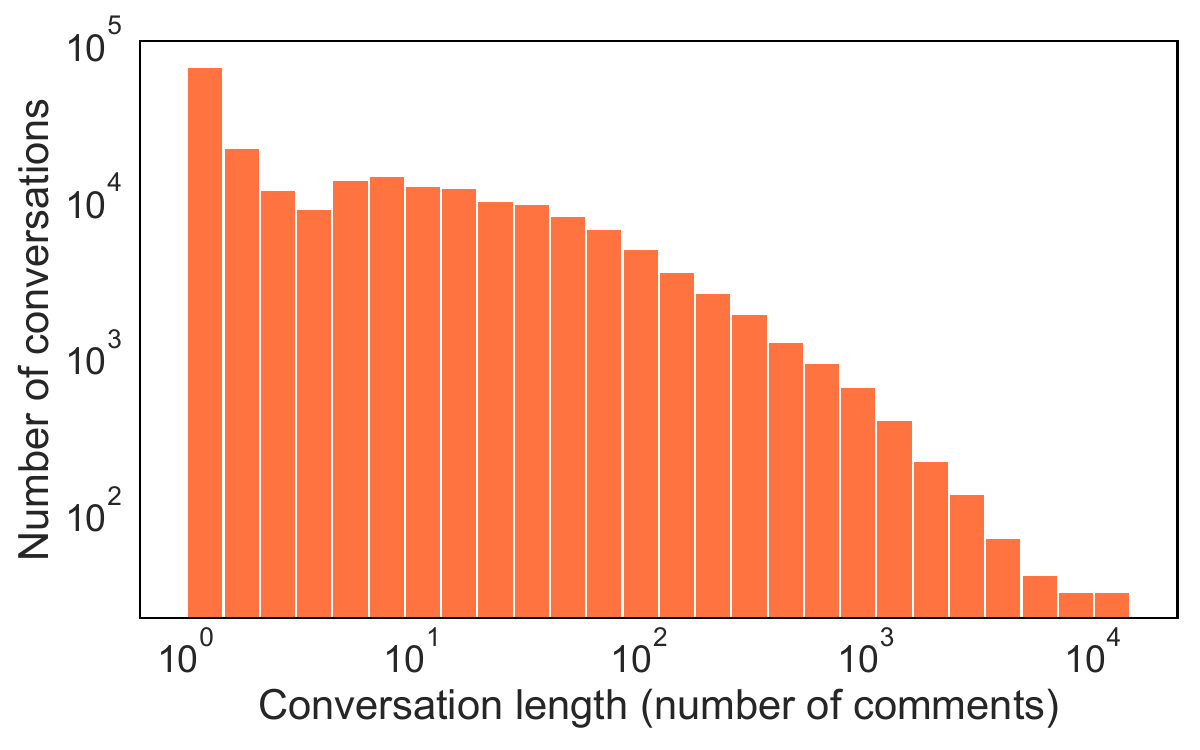}
    \caption{Distribution of conversation lengths in IsamasRed.}
    \label{fig:length_conv}
\end{figure}

\begin{figure*}[tbh]
    \centering
    \includegraphics[width=0.9\textwidth]{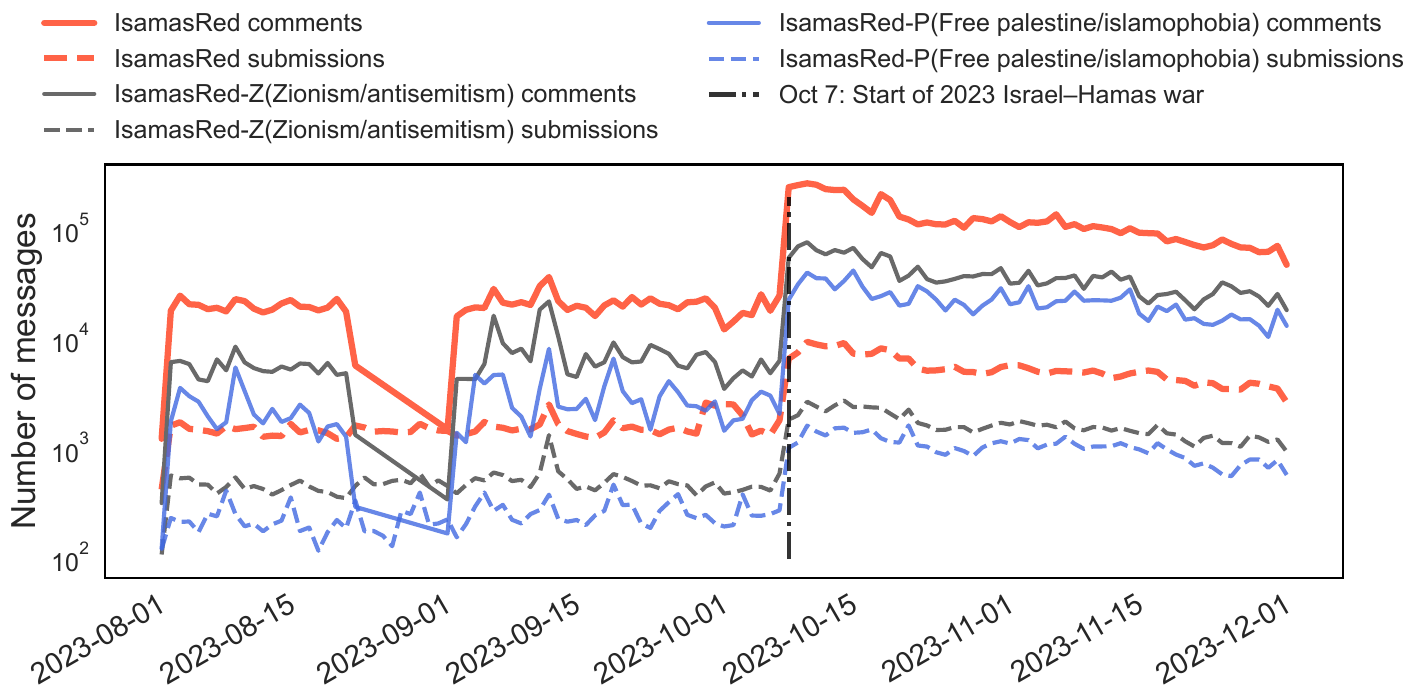}
    \caption{Number of submissions and comments posted over time in IsamasRed, IsamasRed-Z, and IsamasRed-P.}
    \label{fig:n_sub_comm_all}
\end{figure*}

\subsection{Discourse Subsets: Zionism \& Palestinian Rights}
The impact of the Israel-Hamas conflict has reverberated internationally, leading to heated online debates about Zionism, antisemitism, ``Free Palestine'' movement and Islamophobia. Zionism and antisemitism are often intertwined with Israeli and Jewish identities and the perceptions of bias or prejudice. Similarly, the ``Free Palestine'' movement and charges of Islamophobia can reflect stances on Palestinian rights and anti-Muslim sentiments. 
In this regard, we compile two subsets on top of IsamasRed, namely IsamasRed-Z and IsamasRed-P, pertinent to Zionism/antisemitism and Free Palestine/Islamophobia topics respectively. 
We employ our method of automated keyword extraction with LLM again to identify keywords pertinent to these two topics.
Specifically, for Zionism/antisemitism, we use seed terms \emph{Zionism} and \emph{antisemitism} to retrieve the Wikipedia pages, and for Free Palestine/Islamophobia we use \emph{Free Palestine} and \emph{islamophibia}. The extracted keywords for the two topics are listed in the Appendix. With the keywords, we filter the relevant submissions from IsamasRed and curate IsamasRed-Z and IsamasRed-P datasets. The statistics of these two datasets are shown in Table \ref{tab:data_stats}.


\section{Statistical Analysis}
In this section we conduct an initial analysis on the statistics of our dataset.

\subsection{Submissions and Comments}

In the context of the Israel-Hamas conflict, user-generated submissions on Reddit typically encompass news articles, analytical think pieces, opinion editorials, and personal narratives directly related to the evolving situation. These submissions provide updates on the conflict, delve into the international ramifications, and often present historical context. In response, comments become a dynamic forum for further discussion, allowing users to debate varied perspectives, share additional insights, or present contrasting viewpoints. This interactive layer enriches the discourse, enabling a multifaceted exploration of public's reactions to the conflict. 

Figure \ref{fig:n_sub_comm_all} illustrates the distribution of the number of submissions and comments posted over time in IsamasRed, IsamasRed-Z and IsamasRed-P, reflecting user engagement patterns. Following the Hamas attack on October 7, there is a clear exponential growth in both submissions and comments, signifying increased community attention to the conflict. There is a discernible spike in submissions at the beginning of October, indicating the escalating tensions or significant events leading to the conflict.
It is also observed that submissions and comments related to ``Zionism and antisemitism'' are more dominant than those concerning ``Free Palestine and Islamophobia.'' 

\subsection{Conversations}


A Reddit conversation is an interactive exchange that starts with an original submission and evolves through subsequent user comments and replies, forming a structured discussion. Each conservation is associated with a submission.
Conversations about the Israel-Hamas conflict provide insights into public sentiment, evolving viewpoints, and the complexities of community interaction. Reflecting prevailing sentiments and controversies, these discussions explore various aspects of the conflict through a nested, branching structure. As they evolve, early comments may influence the trajectory of the dialogue, with new information and perspectives emerging as more users participate. 
In Figure \ref{fig:length_conv}, we show the distribution of conversation lengths in IsamasRed, which is skewed: the vast majority of conversations have one hundred comments or less, although a few conversations have thousands of comments.

\begin{figure*}[ht]
    \centering
    \includegraphics[width=0.9\textwidth]{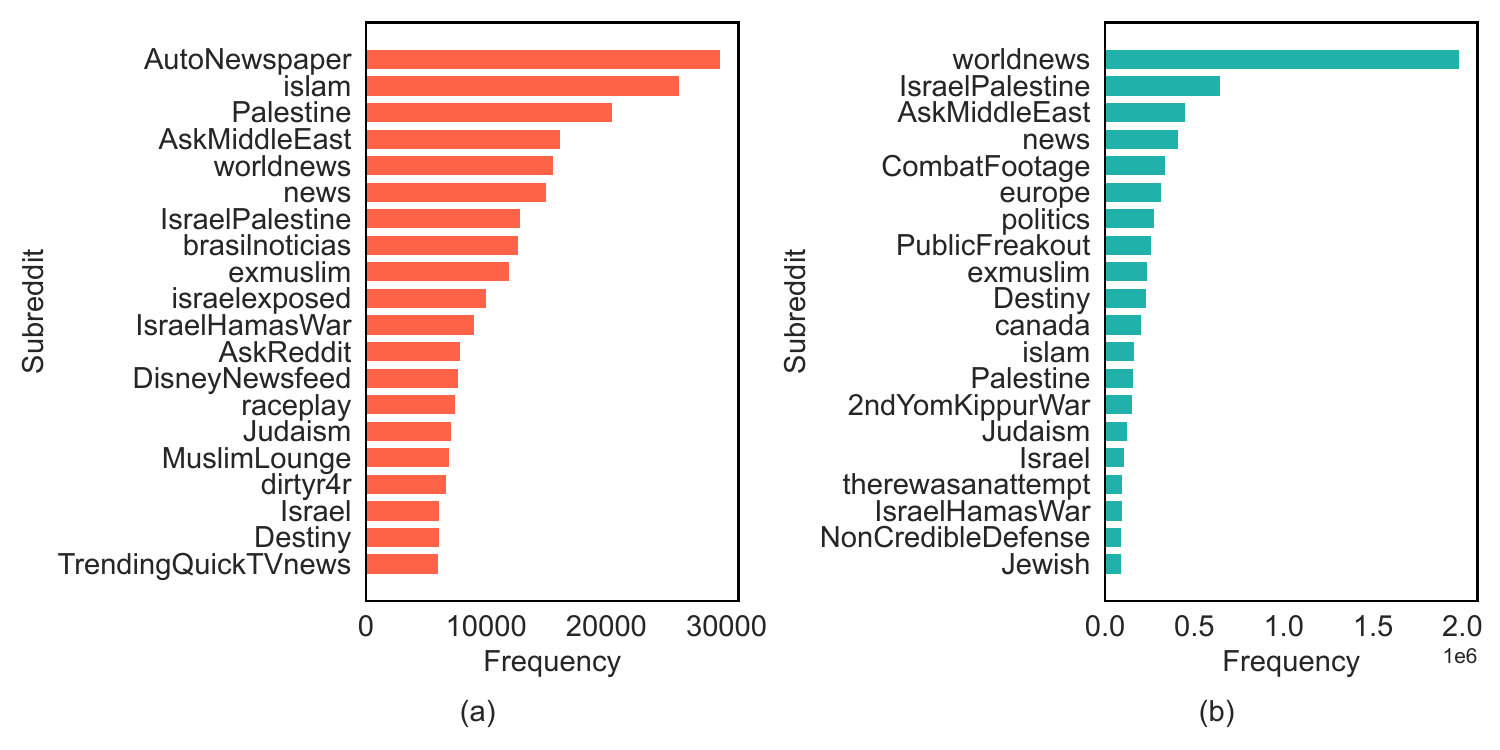}
    \caption{Top 20 largest subreddits by the number of (a) submissions and (b) comments.}
    \label{fig:top_subreddit}
\end{figure*}

\subsection{Subreddits}

Subreddits function as community-driven forums on Reddit, each focusing on a specific topic. For a topic like the Israel-Hamas conflict, a relevant subreddit would serve as a dedicated space where individuals interested in this subject can share and discuss news articles, personal opinions, and questions about the war. 
Figure \ref{fig:top_subreddit} shows the top 20 subreddits based on the volume of posts and comments. Conflict-inclusive subreddits categorized under news tend to exhibit higher frequency, attributable to their larger user base and their higher activity levels. Furthermore, the topical relevance of the Israel-Hamas conflict to these news forums contributes to their prominence in our findings. Additionally, certain conflict-centric subreddits like \emph{r/IsraelPalestine}, \emph{r/AskMiddleEast}, and \emph{r/islam} manifest heightened activity levels due to their direct relevance to the conflict.

\section{Semantic Analysis}
In this section we anaylze the semantics of \textbf{comments} in IsamasRed, including user engagement, controversiality, moral foundations, and emotions.

\subsection{User Endorsement}

\begin{figure}[ht]
    \centering
    \includegraphics[width=0.48\textwidth]{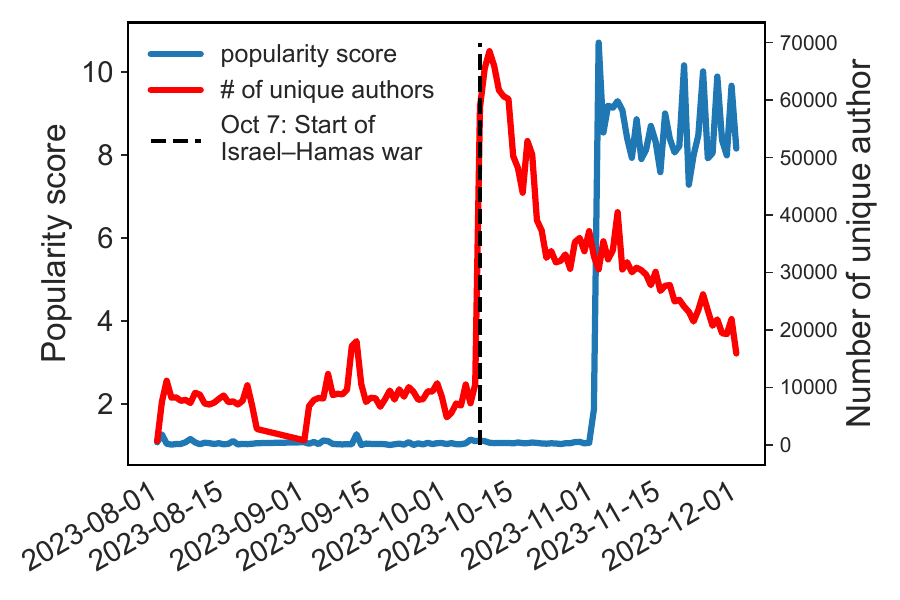}
    \caption{Popularity score and the number of unique authors of comments over time.}
    \label{fig:pop_score}
\end{figure}


User endorsement metrics, such as the popularity score (the number of upvotes minus the number of downvotes), the ratio of upvotes, and total upvotes, offer valuable insights for analyzing the influence and reach of social media discussions surrounding the conflict. We show the popularity scores comments and the number of unique active users over time in Figure \ref{fig:pop_score}.
There is a significant spike in the popularity score immediately following the beginning of November. This spike likely reflects a surge in user engagement and interest in the conflict, resulting in more active voting on related content. However, there is a consistent flat line before November due to the API update strategy. Following the initial peak, the popularity score settles into a pattern of fluctuation, which may indicate ongoing but less intense engagement with the topic as the conflict continues. 

We further verified user engagement in the conflict, as indicated by the daily count of unique authors. There is a pronounced peak around the same time as the beginning of the Israel-Hamas conflict around October 7th, suggesting that the beginning of the conflict motivated a large number of users to contribute to the conversation. This influx of authors could be due to heightened interest, a desire to share information, or to express opinions and reactions to the unfolding events. Following the spike, the number of unique authors shows a more consistent downward trend, suggesting that fewer new contributors are joining the discussion over time.

\subsection{Controversiality}

\begin{figure*}[ht]
    \centering
    \includegraphics[width=1\textwidth]{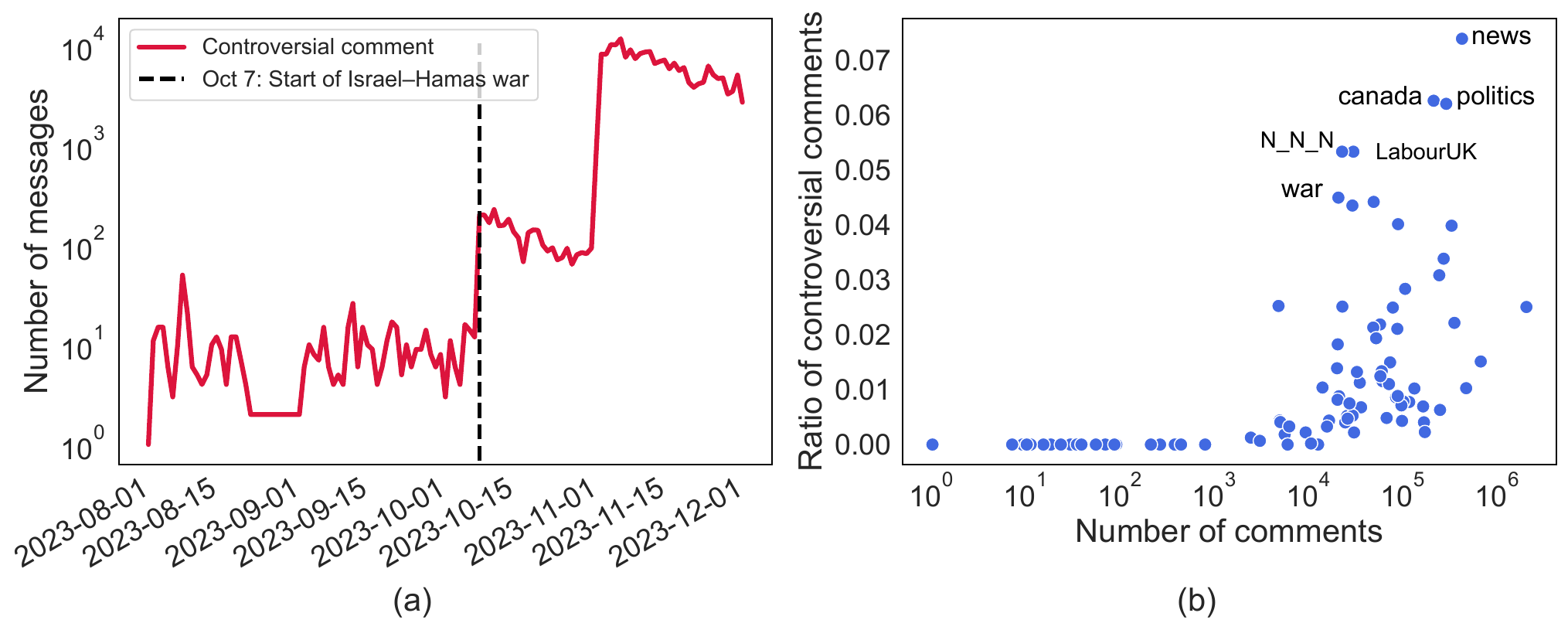}
    \caption{(a) Number of controversial comments over time. (b) Ratio of controversial comments in each subreddit as a function of the total number of comments in it.}
    \label{fig:controvs}
\end{figure*}


The controversy in the Israel-Hamas conflict discourse on Reddit stems from its complex historical and political roots, highly charged emotions \cite{chen2023anger}, diverse global perspectives, and the intricate mix of legitimate critique and prejudice. This mix of factors creates grounds for heated, often contentious, debates among users with deeply held convictions and backgrounds.
IsamasRed includes a binary controversiality indicator for comments, automatically assigned by Reddit to comments that receive a large number of upvotes and downvotes simultaneously. A conservation is marked controversial if it contains at least one controversial comment.
Previous studies have used this indicator to explore controversy on Reddit \cite{koncar2021analysis,chen2023anger}. 

Figure~\ref{fig:controvs}(a) shows the number of controversial comments over time, showing a notable increase on October 7th, the date of the Hamas attack. 
An additional increase in controversiality is seen in early November, which might be attributed to the data collection methodology of Pushshift API. 
Overall, the analysis suggests that the Israel-Hamas conflict has led to an increase in controversial comments, reflecting the intensified discourse and engagement on this topic.

In Figure \ref{fig:controvs}(b), we present the relationship between the ratio of controversial comments and total number of comments across various subreddits. Each dot represents a subreddit.
In general, subreddits with a high volume of comments (e.g. \emph{r/news}) also has a significant number of controversial comments. These subreddits have more participants, leading to increased interactions and voting dynamics. Subreddits with larger numbers of comments, such as \emph{r/worldnews}, show a lower ratio of controversial comments, potentially due to a broader audience, a wider variety of discussions, and strict moderation rules, diluting the concentration of controversy.

\begin{figure*}
    \centering
    \includegraphics[width=0.9\textwidth]{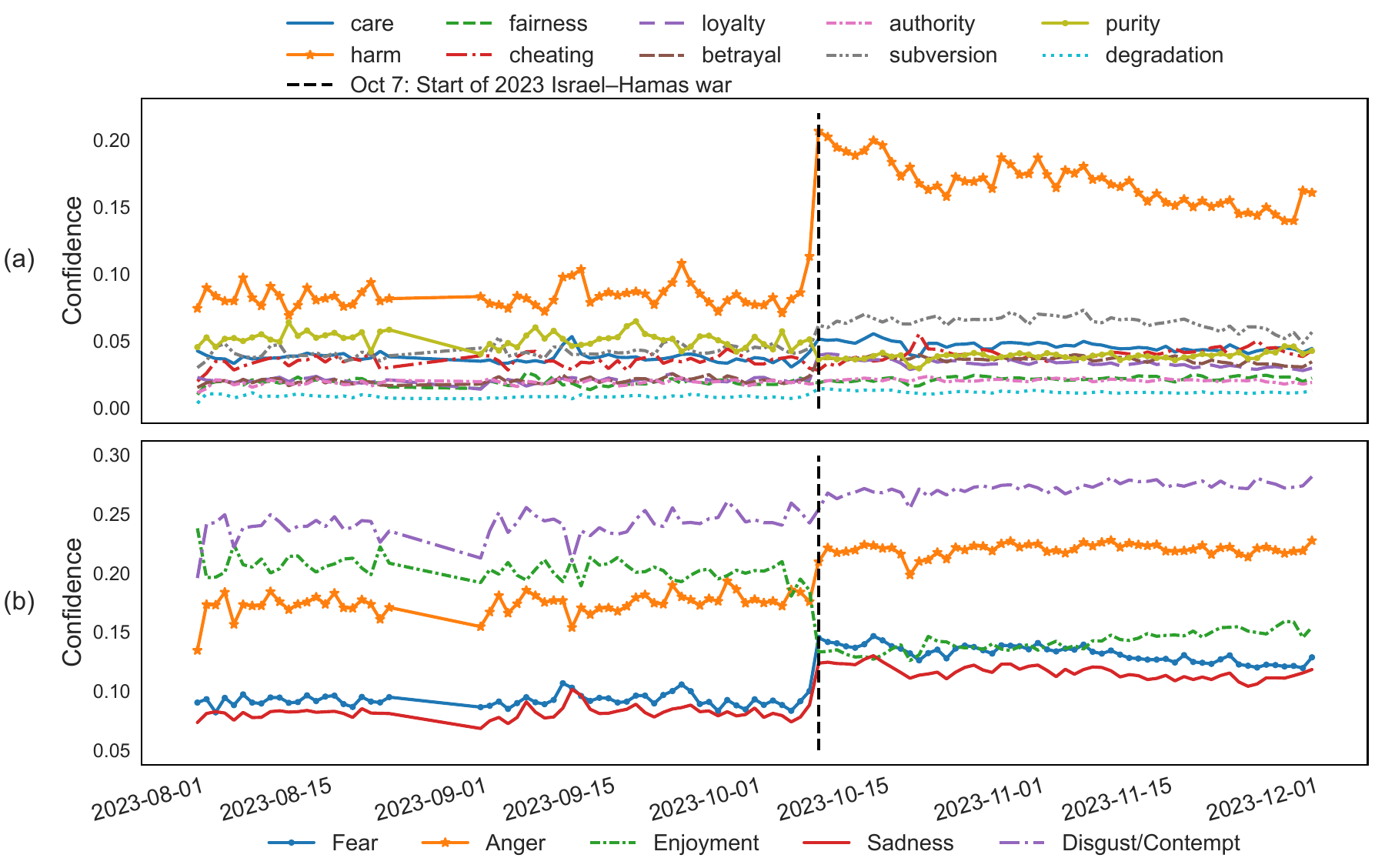}
    \caption{(a) Moral foundations and (b) emotions of comments over time. The average confidence of each moral foundation/emotion in the comments posted each day is shown.
    }
    \label{fig:mf_emotion}
\end{figure*}

\subsection{Moral Foundations}

The moral dimensions of the discussions of war often revolve around the foundations of care (and harm) for human lives, fairness (and cheating), loyalty to (and betrayal of) group or national identity, respect for authority and historical claims (or subversion of these), and the sanctity (or degradation) of religious or ideological beliefs. These moral foundations underpin the culture of diverse groups and drive the ethical and political values that individuals bring to these discussions.

We use the Domain Adaptation Moral Foundation (DAMF) model~\cite{guo2023data} to label the moral dimensions of Reddit comments. DAMF employs a data fusion framework to enhance language models for morality inference across heterogeneous datasets. For each comment it assigns ten distinct binary labels reflecting moral foundations: \emph{care}, \emph{harm}, \emph{fairness}, \emph{cheating}, \emph{loyalty}, \emph{betrayal}, \emph{authority}, \emph{subversion}, \emph{purity}, and \emph{degradation}, as shown in Figure \ref{fig:mf_emotion}(a).

The most prominent foundation is harm, which shows a significant rise around the start of the conflict, indicating a surge in discussions or events related to harm or suffering. This is followed by a notable but less pronounced peak in \textit{care} and \textit{subversion}, indicating an elevated discourse around nurturing, protection, or potential defiance and upheaval in response to the conflict. \textit{Fairness} and \textit{betrayal}, which could suggest a response to perceived injustices or treacheries related to the conflict's events. Meanwhile, \textit{purity} declines post-conflict, suggesting a reduced emphasis on sanctity or corruption within the narrative. Other moral foundations remain relatively stable but at lower confidence levels, indicating these themes were less dominant in the discourse. The overall pattern suggests that the moral rhetoric during this period was dominated by considerations of harm and subversion of authority, which are often central in conflict situations.

\subsection{Emotions}

Emotions are an important driver of user engagement with online communities. Emotions shape perspectives, foster solidarity or division, and impact the quality of discourse. These emotional expressions influence how participants perceive and discuss the conflict, often leading to passionate, polarized debates. Understanding the emotional dimensions of discourse is vital for interpreting the dynamics of the discussions within the dataset. We run an emotion indicator model \cite{chochlakis2023emotion, chochlakis2023leveraging} on the comments in IsamasRed. The model enhances detection by exploiting label correlations and integrating pairwise constraints, achieving improved performance and robustness across multiple languages. For each comment, we measure the following emotions: \textit{fear}, \textit{anger}, \textit{enjoyment}, \textit{sadness}, \textit{disgust/contempt}, and each comment can have multiple emotions.

The emotional analysis of IsamasRed is shown in Figure~\ref{fig:mf_emotion}(b), revealing a predominance of negative emotions, with \textit{anger}, \textit{fear}, and \textit{sadness} markably increasing after the start of the conflict. The positive emotions, such as \emph{enjoyment}, decline. 
\textit{Disgust/contempt} is the most prevalent emotion throughout the period, reflecting the strong negative feelings of Reddit users related to the war's events. The sustained levels of \textit{anger} could also indicate a continuous dialogue or narrative around aggression and blame. The sharp decrease in \textit{enjoyment} at the conflict's outset is predictably low, overshadowed by the negative context of war. In addition, \textit{fear} spikes with the conflict's onset, reflecting increased anxiety, while \textit{sadness} shows persistent grief over the war's duration. \textit{Disgust/contempt}, maintaining a high and steady presence throughout the timeline, shows that there is a considerable sense of revulsion or moral judgment, possibly towards acts of violence, the conditions of the war, or the behaviors of certain actors within it.

\section{Release and Access}

Our dataset and code are publicly available on Github\footnote{\url{https://github.com/kaichen23/israel-hamas}} and Zenodo\footnote{\url{https://doi.org/10.5281/zenodo.10494990}}. The repository contains samples of five hundred submissions, comments, and conversations, along with the full ids. Due to the Reddit privacy policy, please contact the authors for access to the texts in the entire dataset.

\section{Limitations}
First, the keyword extraction adopts an English-centric approach. Therefore, IsamasRed might not encompass sufficient multilingual discourse on the topic, especially the main languages of the Israel-Hamas conflict, i.e., Hebrew and Arabic. 

Second, the keywords are extracted from Wikipedia pages, but they are used for data collection on Reddit. There might be some discrepancy in term usage about the conflict between Wikipedia and Reddit, so some keywords extracted from Wikipedia barely helped in data collection on Reddit.

In addition, the keywords are extracted by GPT-4, and the bias inherent in the model may be propagated into this process\cite{santurkar2023whose, he2024whose}.

Moreover, users may deliberately misspell some words (e.g., \emph{Isnotreal} and \emph{g3nocid3}) when discussing the conflict on Reddit, which cannot be collected by our method.
Moreover, since the keywords are mostly in English, 

Last, the data collection relies on third-party sources, resulting in data deficiency from August 20th to August 30th due to the glitch from the sources, as can be observed from Figure \ref{fig:n_sub_comm_all}, \ref{fig:pop_score}, \ref{fig:controvs}(a), \ref{fig:mf_emotion}.

\section{Acknowledgments}
This project was funded in part by DARPA under contract HR001121C0168.

\bibliography{aaai24}


\section{Ethics Statement}
The dataset presented in this study consists entirely of publicly available information. The data collection process was exempted from full review by the Institutional Review Board (IRB) at the University of Southern California, as it relied solely on anonymized, publicly accessible data. Throughout the data collection, analysis, and publication stages, we rigorously adhered to Reddit's terms of service and privacy policy. In ethical integrity, we complied with data collection guidelines, ensuring strict adherence to Reddit's policies, particularly in terms of privacy protection. One possible negative outcome of our work is privacy concerns, especially individual users are identifiable based on author name attribute. To help preserve anonymity, we encoded user names using a hashing function. 
Furthermore, to ensure responsible usage, we have published only a sample of the dataset and invite researchers interested in the full dataset to contact us for access. The dataset has been used exclusively for research purposes, avoiding any form of commercial exploitation or third-party sharing. Our practices underscore our dedication to responsible and ethical academic research.

\appendix

\section{Appendix}
\label{sec:appendix}

\subsection{Prompt Template for Wikipedia Pages Filtering}
\noindent \texttt{Evaluate the content of the Wikipedia page titled [page\_name] to determine if it is related to [topic]. Consider the page name and the first 100 words of the content. Output only YES if the content is related to the war; otherwise, output NO.}\\
\noindent \texttt{First 100 words: [first\_100\_words]}

\subsection{Prompt Template for Keyword Extraction}
\noindent \texttt{Please analyze the provided text chunk from the Wikipedia page about [topic]. Your task is to extract keywords that are directly related to this topic. Each keyword should be no longer than three tokens. After identifying these keywords, evaluate their importance for the purpose of social media message filtration. Rate each keyword's importance on a scale from 0 to 5, where 0 means the keyword is least important and 5 means it is extremely important for filtering messages. Present the output as a list of keyword-importance pairs, separated by commas. Format each pair with the keyword followed by a colon and its importance rating. For example, ``keyword1: 4, keyword2: 2''. The output should be a continuous string text without any line breaks or bullet points.}\\
\noindent \texttt{Text: [text]}

\subsection{Prompt Template for Keyword Filtering}
\noindent \texttt{Please filter the provided list of keywords based on the following criteria:}\\
\noindent \texttt{1. Exclude any keywords that are names of countries or presidents.}\\
\noindent \texttt{2. Exclude any keywords that are names of news organizations or social media platforms, such as ``TikTok'' or ``BBC''.}\\
\noindent \texttt{3. In the list, if a keyword contains another keyword, remove the longer keyword. For example, if the list includes both ``2023 Israel-Hamas war'' and ``Hamas'', remove ``2023 Israel-Hamas war''.}\\
\noindent \texttt{Your task is to process the list and return a filtered set of keywords that meet these criteria. Please present the filtered keywords in a list format.}\\
\noindent \texttt{Keyword list: [keyword\_list]}

\subsection{Retrieved Wikipedia Pages for IsamasRed}

\textit{2023 Gaza humanitarian crisis, 2023 Hamas attack on Israel, 2023 Israeli blockade of the Gaza Strip, 2023 Israeli invasion of the Gaza Strip, 2023 Israel–Hamas war hostage crisis, 2023 Israel–Hamas war protests, 2023 Israel-Hamas war, 2023 Israel-Lebanon border clashes, Al-Shifa Hospital siege, Al-Shifa Hospital, Anti-Palestinianism during the 2023 Israel-Hamas war, Antisemitism during the 2023 Israel–Hamas war, Attacks on health facilities during the 2023 Israel–Hamas war, Casualties of the 2023 Israel-Hamas war, Denial of atrocities during the 2023 Hamas attack on Israel, Diplomatic impact of the 2023 Israel-Hamas war, Economic impact of the 2023 Israel–Hamas war, Effects of the 2023 Israel–Hamas war, Evacuation of the northern Gaza Strip, Evacuations during the 2023 Israel-Hamas war, Francesca Albanese, Gaza City, Gaza Health Ministry, Gaza War, Gaza-Israel conflict, Governance of the Gaza Strip, History of Gaza, Houthi involvement in the 2023 Israel-Hamas war, International reactions to the 2023 Israel-Hamas war, Islamophobia during the 2023 Israel-Hamas war, Israeli war cabinet, Killing of journalists in the 2023 Israel-Hamas war, Killing of Shani Louk, List of engagements during the 2023 Israel-Hamas war, List of journalists killed in the 2023 Israel-Hamas war, March for Israel, Mass detentions in the 2023 Israel-Hamas war, Media coverage of the 2023 Israel-Hamas war, Misinformation in the 2023 Israel-Hamas war, Nukhba (Hamas), Outline of the 2023 Israel-Hamas war, Palestinian genocide accusation, Siege of Gaza City, Timeline of the 2023 Israel–Hamas war, Turkish support for Hamas, United Nations General Assembly Resolution ES-10$\_$21, United States support for Israel in the 2023 Israel-Hamas war, Use of human shields by Hamas, Violent incidents in reaction to the 2023 Israel-Hamas war, War crimes in the 2023 Israel-Hamas war}

\subsection{Keywords for IsamasRed}
\textit{Hamas, Gaza, Israel, IDF, West Bank, war crime, Human Rights Watch, Hezbollah, Amnesty International, Al-Shifa Hospital, United Nations, hostages, Jabalia refugee camp, ceasefire, UNRWA, airstrikes, collective punishment, invasion, Antony Blinken, Shin Bet, international humanitarian law, Operation Al-Aqsa Flood, UNICEF, International Criminal Court, Yoav Gallant, genocide, Ismail Haniyeh, World Health Organization, blockade, Operation Protective Edge, Doctors Without Borders, Al-Shifa hospital, Tel Aviv, Fatah, human shields, Mahmoud Abbas, Rafah Border Crossing, crimes against humanity, Al-Aqsa Mosque, humanitarian crisis, Re'im music festival, Islamophobia, two-state solution, Denial of atrocities, Al Jazeera, UN Security Council, Committee to Protect Journalists, Rafah crossing, Al-Ahli Arab Hospital, Indonesia Hospital, al-Shifa Hospital, Rafah Crossing, antisemitism, journalists killed, Iron Dome, Church of Saint Porphyrius, international law, Mossad, Operation Cast Lead, Recep Tayyip Erdoğan, al-Ahli Arab Hospital, Al-Quds Hospital, Saudi Arabia, humanitarian truce, rocket, aerial bombardment, Antisemitism, Netanyahu, Casualties, Recep Tayyip Erdogan, humanitarian aid, ethnic cleansing, medical supplies, emergency unity government, USS Gerald R. Ford, Oxfam, Qassam Brigades, Beit Hanoun, Middle East, Red Cross, East Jerusalem, Six-Day War, First Intifada, demonstrations, Justin Trudeau, Gilad Shalit, UN Human Rights Council, Death to Arabs, Fourth Geneva Convention, Indonesian Hospital, Oslo Accords, humanitarian pauses, terrorist attack, Islamic Jihad, Eilat, sexual violence, white phosphorus, Antonio Guterres, Operation Pillar of Defense, ICRC, Palestine, Ghazi Hamad, Abdul-Malik al-Houthi, Second Intifada, Operation Swords of Iron, USS Carney, Bureij, Daniel Hagari, terrorist organization, Ceasefire, Rishi Sunak, Olaf Scholz, COGAT, right to self-defense, humanitarian corridor, Khan Yunis, human rights violations, Lebanon, Nuseirat refugee camp, propaganda, Al-Shifa ambulance airstrike, Al-Rantisi Hospital, Be'eri, Palestinian, Casualties, militant groups, al-Quds Hospital, Salah al-Din Road, violence against journalists, ambulance convoy, al-Shifa hospital, Rashida Tlaib, Al-Shati refugee camp, Omar Daraghmeh, Nukhba, Abdel Fattah el-Sisi, tunnel entrance, tunnel network, prisoner exchanges, starvation, medical facilities, state of war, Yom Kippur War, Operation Summer Rains, Khaled Mashal, solar panels, medical neutrality, Yahya Sinwar, Red Sea, Francesca Albanese, apartheid, United States support, hate crime, International reactions, refugee camps, drone strikes, mosques, Sderot, Al-Quds Brigades, al-Maghazi refugee camp, terrorism, Disinformation, military aid, crime against humanity, weapons storage, drones, evacuation, Mohammed bin Salman, anti-tank missiles, Houthi movement, Nakba, massacre, El Hamma synagogue, administrative detention}

\subsection{Conflict-inclusive Subreddits}
\textit{AutoNewspaper, worldnews, news, brasilnoticias, AskReddit, Destiny, 2ndYomKippurWar, CombatFootage, DisneyNewsfeed, TrendingQuickTVnews, Conservative, BreakingNews24hr, conspiracy, EndlessWar, PublicFreakout, politics, NoStupidQuestions, SeenOnNews\_longtail, NBCauto, raceplay, FreeKarma4You, europe, NoFilterNews, worldnewsvideo, TheDeprogram, Mexico\_Videos, dirtyr4r, FRANCE24auto, Ernesto\_it, honestheadlinenews, conservatives, N\_N\_N, Judaism, socialism, Hasan\_Piker, TrendsNewsWorld, NewsWhatever, ItaliaBox, VaushV, theworldnews, TopMindsOfReddit, TIMESINDIAauto, NonCredibleDefense, rustjob, CNNauto, explainlikeimfive, NewsOfTheStupid, ReactJSJobs, anime\_titties, therewasanattempt, ukpolitics, lebanon, ALJAZEERAauto, NYTauto, BBCauto, golangjob, TWTauto, geopolitics, h3h3productions, redscarepod, GUARDIANauto, TheMajorityReport, worldpolitics2, FOXauto, war, NewIran, LabourUK, canada, JavaScriptJob, telaviv, Britain, india, neoliberal, chomsky, infomoney}

\subsection{Retrieved Wikipedia Pages for IsamasRed-Z}
\textit{Jews Against Zionism, Neo-Zionism, Neturei Karta, New antisemitism, Non-Zionism, Politics of Israel, Proto-Zionism, Soviet anti-Zionism, United Nations General Assembly Resolution 3379, Zionism (disambiguation), Zionism as settler colonialism, Zionism, Zionist antisemitism, Anti-antisemitism, Anti-Zionism, Antisemitism during the 2023 Israel–Hamas war, Antisemitism, Claudine Gay, Eliminationist antisemitism, Liz Magill, Self-hating Jew, StopAntisemitism, Timeline of antisemitism
}

\subsection{Keywords for IsamasRed-Z}
\textit{Likud, Neturei Karta, Six-Day War, Jewish, ethnic cleansing, West Bank, Zion, Balfour Declaration, apartheid, Jews, aliyah, Jerusalem, Torah, Palestine, Edah HaChareidis, Satmar, colonialism, religious ideology, National Religious Party, Anti-Defamation League, Yasser Arafat, Arthur Balfour, Theodor Herzl, Abba Eban, kibbutzim, Gush Emunim, national liberation movements, Holocaust, Double standards, Chaim Herzog, Online hate speech, anti-Semitic stereotypes, UN General Assembly Resolution 46/86, homeland, Benjamin Netanyahu, Daniel Patrick Moynihan, Levant, Eliezer Ben-Yehuda, antisemitic, Holocaust, genocide, pogrom, Anti-Defamation League, forced conversion, Ku Klux Klan, persecution, neo-Nazi, blood libel, synagogue destruction, racism, Martin Luther, ghettos, Hamas-led attack, Marranos, usury, Spanish Inquisition, Theodor Herzl, Pale of Settlement, Islamophobia, antisemite, Nazi propaganda, Dreyfus Affair, National Socialist, blood-libels, Einsatzgruppen, Nuremberg Laws, Kristallnacht, far-right political parties, Semitism, hate crime, inferior race, eugenics, Semitic race, Aryan superiority, Nazism, synagogue destroyed, mass burning, Host desecration, expulsion, Forced conversion}

\subsection{Retrieved Wikipedia Pages for IsamasRed-P}
\textit{2023 Israel–Hamas war protests in the United States, 2023 Israel–Hamas war protests, Free Gaza Movement, Free Palestine Movement, From the river to the sea, Harbu Darbu, Iran–Palestine relations, Israeli–Palestinian conflict, List of supporters of the BDS movement, Marc Lamont Hill, Outline of the 2023 Israel–Hamas war, Palestine Action, Palestine, Palestinian genocide accusation, Palestinian liberation (disambiguation), Palestinian nationalism, Paul Larudee, Refaat Alareer, Students for Justice in Palestine, United Nations Partition Plan for Palestine, Amira Elghawaby, Collective Against Islamophobia in France, Cultural racism, International Day To Combat Islamophobia, Islamophobia during the 2023 Israel–Hamas war, Islamophobia in the media, Islamophobia, Islamophobia\_Islamophilia
}

\subsection{Keywords for IsamasRed-P}
\textit{genocide, Palestinian, Second Intifada, First Intifada, Jerusalem, Palestine, self-determination, war crimes, Balfour Declaration, Al-Aqsa Mosque, Jewish, International Solidarity Movement, ethnic cleansing, antisemitism, Mahmoud Abbas, apartheid, Benjamin Netanyahu, free speech, Jordan River, human rights, antisemitic, Arab League, BDS movement, Six-Day War, occupation, one-state solution, right of return, Anti-Defamation League, Tel Aviv, Holocaust survivors, Amnesty International, Jews, terrorism, UN General Assembly, peace process, Elbit Systems, International Day of Solidarity, Oslo Accords, Bethlehem, Antony Blinken, Arafat, Al-Aqsa clashes, Hebron, Islamophobia, fear of Islam, Islamophobic, discrimination, discriminatory practices, terrorism, terrorist attacks, September 11, prejudice against Islam, violent Islam, Islamist, racism, multiculturalism, barbaric, monolithic bloc, Islamic, fundamentalism, Islamophilia, CAIR, xenophobia, Sharia law, Islam Awareness Week, English Defence League, sexist, primitive, clash of civilizations, hijab, 9/11, physical attacks, Charlie Hebdo, random assaults, potential terrorists, Muslim, oppression, extermination, Christchurch mosque shootings, critical race theory, identity politics, war on terror, Orientalism, surveillance of mosques, Death to Arabs, anti-semitism, hate crime, media portrayals}

\end{document}